# Practical Denoising of MEG Data Using Wavelet Transform

Abhisek Ukil

Tshwane University of Technology, Pretoria, 0001, South Africa
abhiukil@yahoo.com

**Abstract.** Magnetoencephalography (MEG) is an important noninvasive, non-hazardous technology for functional brain mapping, measuring the magnetic fields due to the intracellular neuronal current flow in the brain. However, the inherent level of noise in the data collection process is large enough to obscure the signal(s) of interest most often. In this paper, a practical denoising technique based on the wavelet transform and the multiresolution signal decomposition technique is presented. The proposed technique is substantiated by the application results using three different mother wavelets on the recorded MEG signal.

## 1   Introduction

Magnetoencephalography (MEG) is completely noninvasive, non-hazardous technology for functional brain mapping. Every current generates a magnetic field, and following this same principle in the nervous system, the longitudinal neuronal current flow generates an associated magnetic field. MEG measures the intercellular currents of the neurons in the brain giving a direct information on the brain activity, spontaneously or to a given stimulus. That is, MEG detects weak extracranial magnetic fields in the brain, and allows determination of their intracranial sources [1].

Unlike Computed Tomography (CT) or Magnetic Resonance Imaging (MRI), which provide structural/anatomical information, MEG provides functional mapping information. By measuring these magnetic fields, scientists can accurately pinpoint the location of the cells that produce each field. In this way, they can identify zones of the brain that are producing abnormal signals. These spatiotemporal signals are used to study human cognition and, in clinical settings, for preoperative functional brain mapping, epilepsy diagnosis and the like.

One common method of collecting functional data involves the presentation of a stimulus to a subject. However, most often the inherent noise level in the data collection process is large enough to obscure the signal(s) of interest. In order to reduce the level of noise the stimulus is repeated for as many as 100-500 trials, the trials are temporally aligned based on the timing of the stimulus presentation, and then an average is computed. This ubiquitously-used approach works well, but it requires numerous trials. This in turn causes subject fatigue and, therefore, limits the number of conditions that can be tested for a given subject.

In this paper, a practical denoising technique of the MEG data using the wavelet transform is presented with application results. The remainder of the paper is organized as follows. In Section 2, practical MEG technique and the associated noise problem is discussed in details. Section 3 provides a brief review of the wavelet transform. Section 4 discusses about the denoising technique using the wavelet transform along with the application results, and conclusion is given in Section 5.

## 2   MEG Technique and Noise Problem

MEG technique measures the extremely weak magnetic field (of the order of femto Tesla, 1 fT = $10^{-15}$ Tesla) generated by the intracellular neuronal current flow in the brain. This was initiated by the first recordings of the human magnetic alpha rhythm by Cohen in 1968 [2].

The spontaneous or evoked magnetic fields emanating from the brain induce a current in some induction coils, which in turn produce a magnetic field in a special device called a superconducting quantum interference device (SQUID) [3]. The MEG sensors consist of a flux transformer coupled to a SQUID, which amplifies the weak extracranial magnetic field and transforms it into a voltage. Present-day whole-head MEG devices typically contain 64-306 sensors for clinical and experimental works. Overall, MEG technique provides high resolution measurement both in space (2-3 mm) and time (1 ms).

Different techniques have been proposed for analysis of the noisy MEG signals, like, independent component analysis [4], maximum-likelihood technique [5], blind source separation [6] etc. In this paper, we present the wavelet transform-based practical denoising technique of the MEG signals.

The experimental setup used in this work consisted of 274 sensors detecting the magnetic field (fT) for pre- and post-stimulus period, while the stimulus is presented to the subject at time $t = 0$ ms. The total duration of the recording of the sensor data for each trial is for 361 ms, of which 120 ms is for pre- and 241 ms is for post-stimulus period. We are interested for the analysis of the post-stimulus period. 10 trials of the MEG recorded signals using the above-mentioned experimental setup have been used for the experimentation.

## 3   Wavelet Transform

The Wavelet transform (WT) is a mathematical tool, like the Fourier transform for signal analysis. A wavelet is an oscillatory waveform of effectively limited duration that has an average value of zero. Fourier analysis consists of breaking up a signal into sine waves of various frequencies. Similarly, wavelet analysis is the breaking up of a signal into shifted and scaled versions of the original (or mother) wavelet. While detail mathematical descriptions of WT can be referred to in [7], [8], a brief mathematical summary of WT is provided in the following sections.

The continuous wavelet transform (CWT) is defined as the sum over all time of the signal multiplied by scaled and shifted versions of the wavelet function $\psi$. The CWT of a signal $x(t)$ is defined as

$$CWT(a,b) = \int_{-\infty}^{\infty} x(t)\psi_{a,b}^{*}(t)dt ,  \quad (1)$$

where

$$\psi_{a,b}(t) = |a|^{-1/2} \psi((t-b)/a) . \quad (2)$$

$\psi(t)$ is the *mother* wavelet, the asterisk in (1) denotes a complex conjugate, and $a, b \in R, a \neq 0$, ($R$ is a real continuous number system) are the *scaling* and *shifting* parameters respectively. $|a|^{-1/2}$ is the normalization value of $\psi_{a,b}(t)$ so that if $\psi(t)$ has a unit length, then its scaled version $\psi_{a,b}(t)$ also has a unit length.

Instead of continuous scaling and shifting, the mother wavelet maybe scaled and shifted discretely by choosing $a = a_0^m, b = na_0^m b_0, t = kT$ in (1) & (2), where $T = 1.0$ and $k, m, n \in Z$, ($Z$ is the set of positive integers). Then, the discrete wavelet transform (DWT) is given by

$$DWT(m,n) = a_0^{-m/2} \left( \sum x[k]\psi^{*}[(k - na_0^m b_0)/a_0^m] \right). \quad (3)$$

By careful selection of $a_0$ and $b_0$, the family of scaled and shifted mother wavelets constitutes an orthonormal basis. With this choice of $a_0$ and $b_0$, there exists a novel algorithm, known as *multiresolution signal decomposition* [9] technique, to decompose a signal into scales with different time and frequency resolution. The MSD [9] technique decomposes a given signal into its detailed and smoothed versions. MSD technique can be realized with the cascaded *Quadrature Mirror Filter* (QMF) [10] banks. A QMF pair consists of two finite impulse response filters, one being a low-pass filter (LPF) and the other a high-pass filter (HPF).

## 4 Denoising Using Wavelet Transform

For denoising purpose, first all the 274 sensor recordings (for the post-stimulus period) are concatenated as a single vector of size 1x66034 (66034=274x241). This is followed by denoising using the wavelet transform. MSD [9] approach is used, for 8 scales, using different mother wavelets. This results in ($2^8 =$)256 times less samples. So, we get an estimate for 66034/256= 258 sensor data. For the rest of the sensors, i.e. 274-258=16, are estimated from the recordings as the mean. These are concatenated with the estimated 256 data from the wavelet analysis to get the 274 sensor data estimation. The final output variable (denoised MEG signal) is constructed by iterating for the 241 post-stimulus period using the denoised estimation. This approach can be applied to get the denoised signal for single representative trial, or for *n* number of trials (iteratively) followed by the average. Obviously the single trial estimation is faster, but the *n*-trial estimation results in better signal quality. If the MSD *N*-scale decomposition results in less number of sensor data (like the case here), we have to perform end-point signal estimation; otherwise if the decomposition results in more number of sensor data, we have to throw away the end-points. We have used three different mother wavelets, Daubechies 4 [7], Coiflets [7] and Adjusted Haar [11]. Fig. 1 shows the average MEG data for the post-stimulus period.

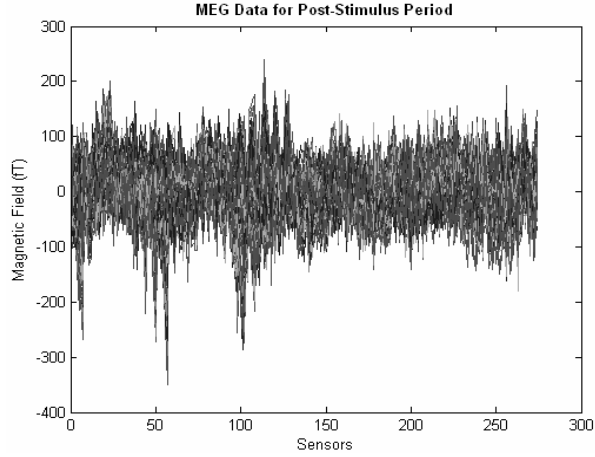

**Fig. 1.** Average MEG Signal over 10 Trials for 274 Sensors

### 4.1 Analysis Using Daubechies 4 Mother Wavelet

For the Daubechies 4 [7] wavelet, the scaling function $\phi(x)$ has the form

$$\phi(x) = c_0\phi(2x) + c_1\phi(2x-1) + c_2\phi(2x-2) + c_3\phi(2x-3) \tag{4}$$

where

$$c_0 = (1+\sqrt{3})/4,\ c_1 = (3+\sqrt{3})/4,\ c_2 = (3-\sqrt{3})/4,\ c_4 = (1-\sqrt{3})/4. \tag{5}$$

The Daubechies 4 wavelet function $\psi(x)$ for the four-coefficient scaling function is given by

$$\psi(x) = -c_3\phi(2x) + c_2\phi(2x-1) - c_1\phi(2x-2) + c_0\phi(2x-3). \tag{6}$$

We used the daubechies 4 (db4) mother wavelet for the 8-scale signal decomposition and denoising. The end-point estimation was done by iterating over the 10 trials. Fig. 2 shows the denoised signal using the db4 mother wavelet compared against the noisy signal in Fig. 1. The magnitude of the magnetic field (Y-axis) remains more or less at same scale while reducing the superimposed noisy components.

### 4.2 Analysis Using Coiflet 1 Mother Wavelet

Coiflets are compactly supported symmetrical wavelets [7]. It has orthonormal wavelet bases with vanishing moments not only for the wavelet function $\psi$, but also for the scaling function $\phi$. For coiflets, the goal is to find $\psi$, $\phi$ so that

$$\int x^l \psi(x)dx = 0, \quad l = 0,1,...,L-1 \tag{7}$$

$$\int \phi(x)dx = 1, \quad \int x^l \phi(x)dx = 0, \quad l = 0,1,...,L-1. \tag{8}$$

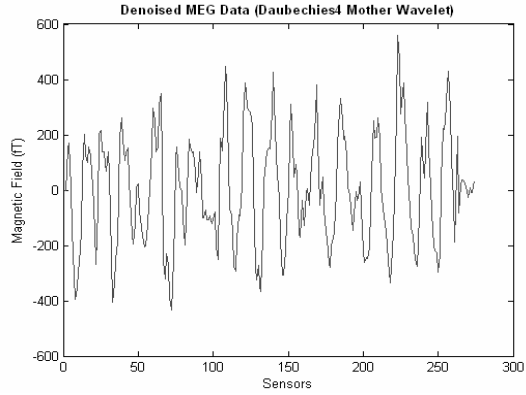

**Fig. 2.** Denoised MEG Signal using the Daubechies 4 Mother Wavelet

*L* is called the *order* of the coiflet [7]. Following several tests, we have chosen *L*=1 for our application, which provided the best denoising performance. The 8-scale signal denoising is followed by the end-point estimation by iterating over the 10 trials. Fig. 3 shows the denoised signal using the coiflet 1 mother wavelet.

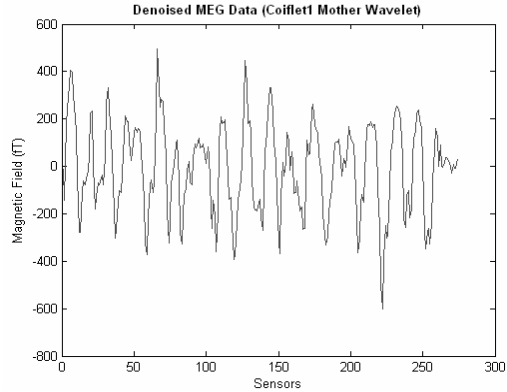

**Fig. 3.** Denoised MEG Signal using the Coiflet 1 Mother Wavelet

### 4.3 Analysis Using Adjusted Haar Mother Wavelet

In general, the FIR (finite impulse response) scaling filter for the Haar wavelet is $h = 0.5[1 \quad 1]$, where 0.5 is the normalization factor. As an adjustment and improvement of the characteristics of the Haar wavelet, Ukil & Zivavovic proposed to introduce 2n zeroes (n is a positive integer) in the Haar wavelet scaling filter, keeping the first and last coefficients 1 [11]. The scaling filter kernel for the adjustment parameter n is shown below.

$$\begin{aligned} h &= 0.5[1 \quad 1] & \text{for } n = 0 \\ h &= 0.5[1 \ 0 \ 0 \ 1] & \text{for } n = 1 \\ h &= 0.5[1 \ 0 \ 0 \ 0 \ 0 \ 1] & \text{for } n = 2 \end{aligned} \quad (9)$$

It should to be noted that the original Haar wavelet scaling filter corresponds to $n = 0$, and complex conjugate pairs of zeroes for each $n > 0$ are introduced [11].

It has been shown mathematically in [11] that the introduction of the adjusting zeroes does not violate the key wavelet properties like compact support, orthogonality and perfect reconstruction. A theorem has been proven in [11] which states:
"The introduction of the 2n adjusting zeroes to the Haar wavelet scaling filter improves the frequency characteristics of the adjusted wavelet function by an order of 2n+1."

Following the proof, the adjusted wavelet function $\psi_n(\omega)$ of the adjusted Haar wavelet becomes,

$$|\psi_n(\omega)| = \frac{\{\sin((2n+1)\omega/4)\}^2}{|(2n+1)\omega/4|} < \frac{4}{|(2n+1)\omega|} . \quad (10)$$

The factor 2n+1 in the denominator of (18) improves the frequency characteristics of the adjusted Haar wavelet function, by decreasing the ripples (as $n > 0$) [11].

We used the adjusted Haar mother wavelet with 4 adjusting zeros for the 8-scale signal denoising. Four zeros were chosen for best possible performance without hampering the speed. Fig. 4 shows the denoised signal using the adjusted Haar wavelet.

### 4.4 Performance

The performance metric used is the signal-to-interference/noise ratio,

$$Output\ SNIR \cong 10 \log_{10} \left( \frac{1}{K} \sum_{i=1}^{K} \frac{\sum_{n=1}^{N} Y_{mean}^2}{\sum_{n=1}^{N} (Y_{mean} - Y_{calc})^2} \right) (dB), \quad (11)$$

where $N = 241$ (post-stimulus period), $K = 274$ (no. of sensors), $Y_{mean}$ is the average MEG signal computed over the 10 trials (shown in Fig. 1), and $Y_{calc}$ is the denoised MEG signal using the three different mother wavelets. The output SNIR, indicated as dB, for the denoising operation using the daubechies 4, coiflet 1 and adjusted Haar mother wavelets are -28 dB, -30 dB and -26 dB respectively. Higher values of the output SNIR indicate better performance. Hence, the denoising operation using the adjusted Haar mother wavelet performs best followed by the daubechies 4 and the coiflet 1 mother wavelets.

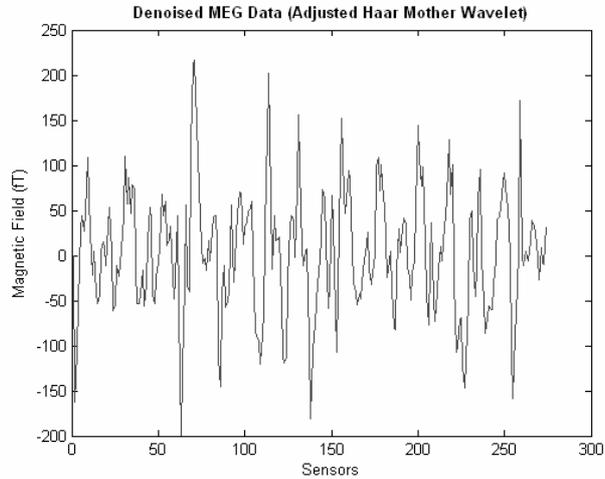

**Fig. 4.** Denoised MEG Signal using the Adjusted Haar Mother Wavelet

The average computation time using the MATLAB® Wavelet toolbox in an Intel® Celeron® 1.9 GHz, 256 MB RAM notebook was 13.42 s, 14.85 s and 13.64 s respectively for the daubechies 4, coiflet 1 and adjusted Haar mother wavelets.

## 5 Conclusion

MEG, the noninvasive technique to measure the magnetic fields resulting from intracellular neuronal current flow, is quite important for functional brain imaging. However, the level of noise that is inherent in the data collection process is large enough that it oftentimes obscures the signal(s) of interest. Normal averaging over numerous trials of signal recording most often does not produce optimum result and also causes subject fatigue. In this paper, we have presented the wavelet transform-based denoising technique of the MEG signal. The concatenated MEG signal from 274 sensors is denoised using the mutiresolution signal decomposition technique. Three different mother wavelets, namely, daubechies 4, coiflet 1 and adjusted Haar have been used for the analysis. The denoising performance is quite robust. Hence, the wavelet tranform-based denoising technique of the MEG signals is quite effective from practical point of view.

## Acknowledgments

The recorded MEG signals were kindly provided by Kenneth E. Hild II, from the Biomagnetic Imaging Lab., Dept. of Radiology, Univ. of California at San Francisco.